%% file: main.tex
\documentclass{article}
\usepackage{amsfonts}
\usepackage{spconf,amsmath,graphicx}
\usepackage{multirow}
\usepackage{subfigure}
\usepackage{booktabs,setspace}
\usepackage{epstopdf}
\usepackage{balance}

\title{THE SLT 2021 CHILDREN SPEECH RECOGNITION CHALLENGE: OPEN DATASETS, RULES and BASELINES}
\name{Fan Yu$^{1*}$\thanks{The first two authors contributed equally to this work.}, Zhuoyuan Yao$^{1*}$, Xiong Wang$^1$, Keyu An$^2$, Lei Xie$^1$, Zhijian Ou$^2$, 
	\emph{Bo Liu$^3$, Xiulin Li$^3$, Guanqiong Miao$^3$}}

\address{$^1$Audio, Speech and Language Processing Group, \\
	School of Computer Science, Northwestern Polytechnical University, Xi'an, China \\
	$^2$Speech Processing and Machine Intelligence Lab, \\
	Tsinghua University, Beijing, China \\
$^3$Databaker Technology, Beijing, China
}
\begin{document}
	%
	\maketitle
	\begin{abstract}
		Automatic speech recognition (ASR) has been significantly advanced with the use of deep learning and big data. However improving robustness, including achieving equally good performance on diverse speakers and accents, is still a challenging problem. In particular, the performance of children speech recognition (CSR) still lags behind due to 1) the speech and language characteristics of children's voice are substantially different from those of adults and 2) sizable open dataset for children speech is still not available in the research community. To address these problems, we launch the Children Speech Recognition Challenge (CSRC), as a flagship satellite event of IEEE SLT 2021 workshop. The challenge will release about 400 hours of Mandarin speech data for registered teams and set up two challenge tracks and provide a common testbed to benchmark the CSR performance. In this paper, we introduce the datasets, rules, evaluation method as well as baselines.
		
	\end{abstract}
	\begin{keywords}
		Automatic speech recognition, children speech recognition, deep learning, audio, datasets
	\end{keywords}
	\input{paper.tex}
	
	\bibliographystyle{IEEEbib}
	\bibliography{strings,refs}
	
\end{document}

%% file: paper.tex
\section{Introduction}
\label{sec:intro}
Automatic speech recognition (ASR) technology has been advanced with the recent help from deep learning (DL)~\cite{lecun2015deep}. A robust speech recognition system is indispensable to a large number of applications in human-computer interaction, online education and audio content analysis. In a broad view, \textit{robustness} means the system has equally good performance for adverse acoustic conditions, diverse speakers and accents, different speaking styles and many other factors. However, current DL-based ASR systems are data-hungry, which is easy to overfit with limited training data. In particular, the speech and language characteristics of children's voices are substantially different from those of adults and ASR for children is still significantly less accurate than that of adults without the coverage of a huge amount of labeled children's speech during training~\cite{russell2007challenges}. Recognizing children's speech has substantial applications, such as computer-assisted language learning (CALL), speech-enabled smart toys, etc.

The difficulty of children's speech recognition lies in the differences in anatomy and language expression between children and adults. Anatomically, the vocal cords of children are shorter and lighter than those of adults. These lead to higher fundamental and resonant frequencies and greater spectral variability of children's speech~\cite{lee1997analysis,lee1999acoustics}. In terms of language expression, children's language skills are obviously worse than those of adults. In particular, children in younger age are unable to express accurately and inevitably have mispronunciations due to limited and evolving language knowledge. Differences also exist between children's and adult's pronunciation in prosody, form, syntax, pragmatics and many other aspects~\cite{potamianos1997automatic}. In addition, another bottleneck to the current development of children speech recognition (CSR) is the lack of sizable labelled children's speech data in the research community.
\begin{table*}[]
    \renewcommand\thetable{2}
	\caption{Challenge track setting.}
	\vspace{5pt}
	\label{tab:track}
	\centering
	\scalebox{1.0}{
		\begin{tabular}{ccc}
			\toprule
			& Track 1                  & Track 2                                                                                               \\ \midrule
			Acoustic model training data & Set A, C1 and C2 & \begin{tabular}[c]{@{}c@{}}Set A, C1, C2 and\\ external data appeared in openslr\end{tabular} \\ \midrule
			Language model training data &
			\begin{tabular}[c]{@{}c@{}}The text transcription in \\ Set A, C1 and C2\end{tabular} &
			\begin{tabular}[c]{@{}c@{}}The text transcription in Set A, C1, C2\\ and external data appeared in openslr\end{tabular} \\ \midrule
			Evaluation data              & \multicolumn{2}{c}{Over 10 hours of Children's reading and conversational speech}                                                                 \\ \bottomrule
	\end{tabular}}
\end{table*}

At the early stage of ASR development, researchers mainly considered the particularity of children's speech and focused on the effect of language models in children's speech recognition. Compared with the adult model, the language model of children's speech is more helpful to children's speech recognition, which showed that there are differences in language expression between children and adults~\cite{das1998improvements}. Some researchers~\cite{Li2002An} also customized specific pronunciation dictionaries for children to improve CSR performance. Fine-tuning the adult language model using child speech data also gained improvements~\cite{gray2014child}. In recent years, with the development of deep learning and the maturity of ASR technology, deep neural network (DNN) provided a great improvement to the acoustic model which has been widely adopted by the research community. DNN can better approach the nonlinear function requested by speech modeling through training with a large amount of data, thus surpassing the ASR system based on Gaussian mixture model (GMM). However, due to the lack of a sizable dataset on children's speech, there are only a few works on DNN to study children's speech recognition. According to~\cite{Giuliani2015LargeVC,Cosi2015A}, training the ASR system based on hybrid DNN-HMM led to slightly improved performance compared with the traditional GMM-based system. There was another method which used a DNN to predict the frequency warping factors for vocal tract length normalization (VTLN) for children, which brought improvement on hybrid DNN-HMM CSR system~\cite{7078563}. Convolutional long short-term memory recurrent neural networks can make great progress on children speech recognition~\cite{Liao2015Large}, and augmenting data by artificially adding noise also achieve stronger robustness. Recent studies~\cite{Liao2015Large,Qian2017Mismatched,Fainberg2016Improving,inproceedings} combined adult speech with children speech for model training while some researchers~\cite{Rong2017Transfer} employed multi-task learning (MTL) which can improve performance for both adults and children. Transfer learning and MTL were also found useful for the CSR task~\cite{DBLP:journals/corr/abs-1809-09658,Gurunath2018Transfer}.

The children Speech Recognition Challenge (CSRC) is a flagship satellite event of IEEE SLT 2021 workshop\footnote{{http://2021.ieeeslt.org/}}. The challenge aims at advancing CSR by releasing a sizable set of labeled children speech data and providing a common testbed to benchmark the CSR performance. Specifically, we will release 400 hours of Mandarin speech data for system building, including 340 hours of adult speech, 30 hours of children reading speech as well as 30 hours of children conversational speech. We particularly set up two challenge tracks: the first track limits the usage of data released by the challenge while the second track allows the participants to use extra constrained data. Performance ranking is based on the character error rate (CER) on the evaluation dataset which is composed of both children reading and conversational speech. Rules and three baselines (Chain model~\cite{Povey2016Purely}, Transformer~\cite{DBLP:journals/corr/VaswaniSPUJGKP17} and CTC-CRF~\cite{CAT_IS20}) are also provided.

The rest of this paper is organized as follows, section 2 will illustrate the released datasets. Section 3 will describe the track setting and evaluation plan. The baseline systems are discussed in Section 4. The challenge rules and other logistics are described in Section 5.

\vspace{-10pt}

\section{Datasets}

Our goal of releasing the open source datasets in this challenge is to ensure the common system building resources and fair evaluation for participants.To this end, we will release single-channel adult speech, child reading and conversational speech for training (totally 400 hours) during the system building stage and additional children's reading and conversational speech for evaluation and ranking in the evaluation stage. The basic information of the training data to be released is summarized in Table~\ref{tab:1}. 
\begin{table}[t]
    \renewcommand\thetable{1}
	\caption{Basic information about the released dataset. A: adult speech training set; C1: children speech training set; C2: children conversation training set.}
	\vspace{5pt}
	\label{tab:1}
	\begin{tabular}{lccc}
		\toprule
		Dataset & A       & C1      & C2             \\ \midrule
		Duration (hrs)  & 341.4   & 28.6    & 29.5           \\
		\# Speakers       & 1999    & 927     & 54             \\
		Speaker age   & 18-60   & 7-11    & 4-11           \\
		Speaking style & Reading & Reading & Conversational \\
		Format & \multicolumn{3}{c}{16kHz, 16bit, single channel wav}     \\
		\bottomrule
	\end{tabular}
\end{table}

\section{TRACK SETTING AND EVALUATION}

We set up two tracks as shown in Table~\ref{tab:track} for participants to study children speech recognition with different limits on the scope of training data. In the first track, participants can only use the data provided by the challenge to train acoustic and language models. In addition, the second track allows the participates to use the data listed in \textit{openslr}\footnote{http://www.openslr.org/}. Note that the language model training in the second track allows the use of transcripts associated with the provided speech data and the external speech data in \textit{openslr}.

There are some differences in the files that need to be submitted between the two tracks. For both tracks, participants need to provide detailed technical solutions and decoding results on the evaluation set, including but not limited to the model structure and algorithms used. Besides, for the second track, participants need to provide a list of additional data used during training.

We use edit distance based CER to evaluate the model performance for both track 1 and 2. The character error rate is calculated as: the summed number of error characters in both the children reading evaluation set and the children conversational evaluation set divided by the total number of characters in the two evaluation sets, as shown in Eq.~(\ref{eq:CER-COMPUTE}):
\begin{equation}\label{eq:CER-COMPUTE}
Score_{CER}= \frac {{CE_{reading}}+{CE_{conversation}}}{{CN_{total}}} \times 100\%,
\end{equation}
where $CE_{reading}$ is the number of character errors on the children reading evaluation set, $CE_{conversation}$ is the number of character errors on the children conversational evaluation set and $CN_{total}$ is the total number of characters in the evaluation set. As standard, insertion, deletion and substitution all account for the errors.

\section{Baseline Systems}
\subsection{Chain-model based hybrid system}
\label{sec:pagestyle}
As a baseline, we implement a chain model based hybrid system using Kaldi~\cite{povey2011kaldi}. The chain model is a type of DNN-HMM model, but there are two differences between chain model and the common DNN model. Firstly, instead of a frame-level objective, chain model uses the log-probability of the phone sequences as the objective function, and computes numerator and denominator under the maximum mutual information (MMI) criterion. Secondly, as for the HMM topology, chain model has a special state accepting phoneme only once and another state accepting silence zero or more times. We briefly review the chain model in the following. Readers can refer to ~\cite{Povey2016Purely} for more details.

The chain model uses the MMI criterion as the objective function. Specifically, in the conventional speech recognition, we only optimize $\sum_{u}P_{\theta}(X|Y)$ in the acoustic model, while  $X = (x_1,x_2,\cdots,x_n)$ is the input feature sequence and $Y = (y_1,y_2,\cdots,y_m)$ is the prediction sequence. But in MMI we hope to optimize the whole formula, and further decompose the denominator with the full probability formula to Eq.~(\ref{eq:mmi}) in which $\theta_{MMI}$ is the objective model parameter.

\begin{equation}
\label{eq:mmi}
\begin{aligned}
\theta_{MMI} &= \mathop{arg max}_{\theta}\sum_{u}log P_{\theta}(Y|X) \\
&= \mathop{arg max}_{\theta}\sum_{u}log \frac{P_{\theta}(X|Y)P(Y)}{P(X)}\\
&= \mathop{arg max}_{\theta}\sum_{u}log \frac{P_{\theta}(X|Y)P(Y)}{\sum_{W}P_{\theta}(X|Y)P(Y)}
\end{aligned}
\end{equation}

Moreover, the chain model uses a model topology different from the conventional HMM which is a 3-state left-to-right HMM that can be traversed in a minimum of 3 frames. The topological structure of chain will output a label representing a phoneme for the first frame of a phoneme, while the output of other frames of the phoneme is similar to blank in the connectionist temporal classification (CTC)~\cite{graves2006connectionist}.

To create the denominator graph, we need a language model and an N-gram model is usually used. Tri-gram and 4-gram are popular choices.

\subsection{Transformer system}	
\begin{figure}[!h]
	\centering
	\includegraphics[scale=0.25]{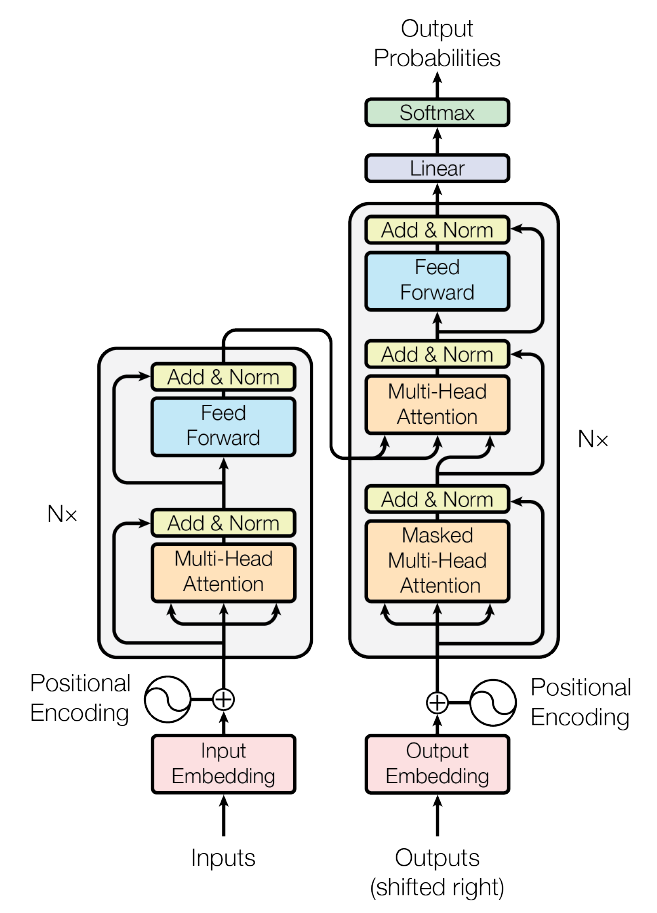}
	\caption{
		Transformer Model~\cite{DBLP:journals/corr/VaswaniSPUJGKP17}
	}
	\label{transformer}
\end{figure}

Besides the conventional hybrid system, we also use a transformer-based end-to-end model as another baseline, which is based on an encoder-decoder structure with an attention mechanism. Transformer was firstly introduced in neural machine translation and achieved superior performance recently in speech recognition~\cite{dong2018speech,li2019speechtransformer,zeyer2019comparison,karita2019comparative}. As shown in Fig.~\ref{transformer}, the encoder can be regarded as a feature extractor that maps the input feature sequence $X = (x_1,x_2,\cdots,x_n)$ to a higher representation sequence $H = (h_1,h_2,\cdots,h_n)$. The decoder decodes $H$ in the way of autoregressive and gets the prediction sequence $Y = (y_1,y_2,\cdots,y_m)$ step by step. The modeling unit of prediction sequence $Y$ needs to be adjusted according to different ASR tasks~\cite{Hu2019Phoneme,Karita2019Improving} in order to obtain a better recognition performance. The commonly used modeling units are phoneme, syllable, character, grapheme or wordpiece.

The encoder and decoder of the transformer are composed of stacked self-attention and position-wise feed-forward layers. In the transformer model multi-head attention plays a significant role, which can obtain information from different representation subspace and each head could focus on a different subspace. Besides, the transformer does not contain the recurrent neural network (RNN) layer, so it cannot obtain any position information of the acoustic sequence. In order to enable the model to make rational use of the location information of the sequence, it is necessary to add the position coding structure to obtain the location information of the corresponding symbols of the whole sequence~\cite{DBLP:journals/corr/VaswaniSPUJGKP17}.

\subsection{CTC-CRF based system}	
The third baseline system is based on the recently proposed CTC-CRF~\cite{CRF_IC20}.
Similar to CTC, CTC-CRF uses a hidden state sequence to obtain the alignment between the label sequence and input feature sequence, and the hidden state sequence is mapped to a unique label sequence by removing consecutive repetitive labels and blanks. The posteriori of the hidden state sequence is defined by a conditional random field (CRF), and the potential function of the CRF is defined as the sum of node potential and edge potential. The definition of node potential is the same as CTC, and the edge potential is realized by an n-gram LM of labels.  By incorporating the probability of the sentence into the potential, the conditional independence between the hidden states in CTC is naturally avoided.

\subsection{Experiment Setup}

We conduct experiments on the released dataset for a total of 400 hours of speech. We manually separate 1\% of the data as a development set and the rest as a training set to benchmark the three baseline systems. We ensure there is no overlap of speakers between the training and development sets. The development set includes adult reading (AR), child reading (CR) and child conversation (CC) subsets. 

In our baseline chain-model system, 71-dimensional mel-filterbanks features are used as the input of the acoustics model and frame length is 25 ms with a 10 ms shift. The acoustic model consists of a single convolutional layer with the kernel of 3 to down-sample the input speech feature, 19 hidden layers of a 256-dimensional Time Delay Neural Network (TDNN) and a 1280-dimensional fully connection layer with activation function of ReLU to obtain the information in time domain and map the speech features to high-dimensional representation. Finally, we use a fully connected layer with ReLU activation function and a log softmax layer to map the feature to the posterior probability of pdf-id. Meanwhile, we train a tri-gram language model using the allowed transcripts and out-of-vocabulary (OOV) words are mapped into $<unk>$. After combining the HMM model, the size of the final  `HCLG` graph is 384M.

We use the ESPnet~\cite{Watanabe2018ESPnet} end-to-end speech processing toolkit to build our transformer baseline. The input features are the same as the chain model.
We follow the standard configuration of the state-of-the-art ESPnet transformer which contains 12-layer encoder and 6-layer decoder, where the dimension of attention and feed-forward layer is set to 256 and 2048 respectively. In all attention sub-layers, 4 heads are used. The whole network is trained for 50 epochs and warm-up~\cite{DBLP:journals/corr/VaswaniSPUJGKP17} is used for the first 25,000 iterations. We use 5140 commonly used Chinese characters as the modeling units. 

We use CAT~\cite{CAT_IS20} to build our CTC-CRF baseline. The AM network is two blocks of VGG layers followed by a 6-layer BLSTM. The first VGG block has 3 input channels corresponding to spectral features, delta, and delta delta features. The BLSTM has 512 hidden units per direction and the total number of parameters is 37M. In training, a dropout probability of 0.5 is applied to the LSTM to prevent overfitting. A CTC loss with a weight 0.01 is combined with the CRF loss to help convergence. We use the open-source Aishell lexicon\footnote{{http://www.openslr.org/resources/33/resource\_aishell.tgz}} as our basic lexicon, and then segment the training transcription, and add the OOV (out-of-vocabulary) words to the lexicon. The language model is a tri-gram word based language model, which is trained on the training transcription after segmentation. The size of decoding graph is 142M.

\subsection{Results and Analysis}
\begin{figure}[!h]
	\centering
	\subfigure{
		\begin{minipage}{6cm}
		    \centering
		    \includegraphics[scale=0.5]{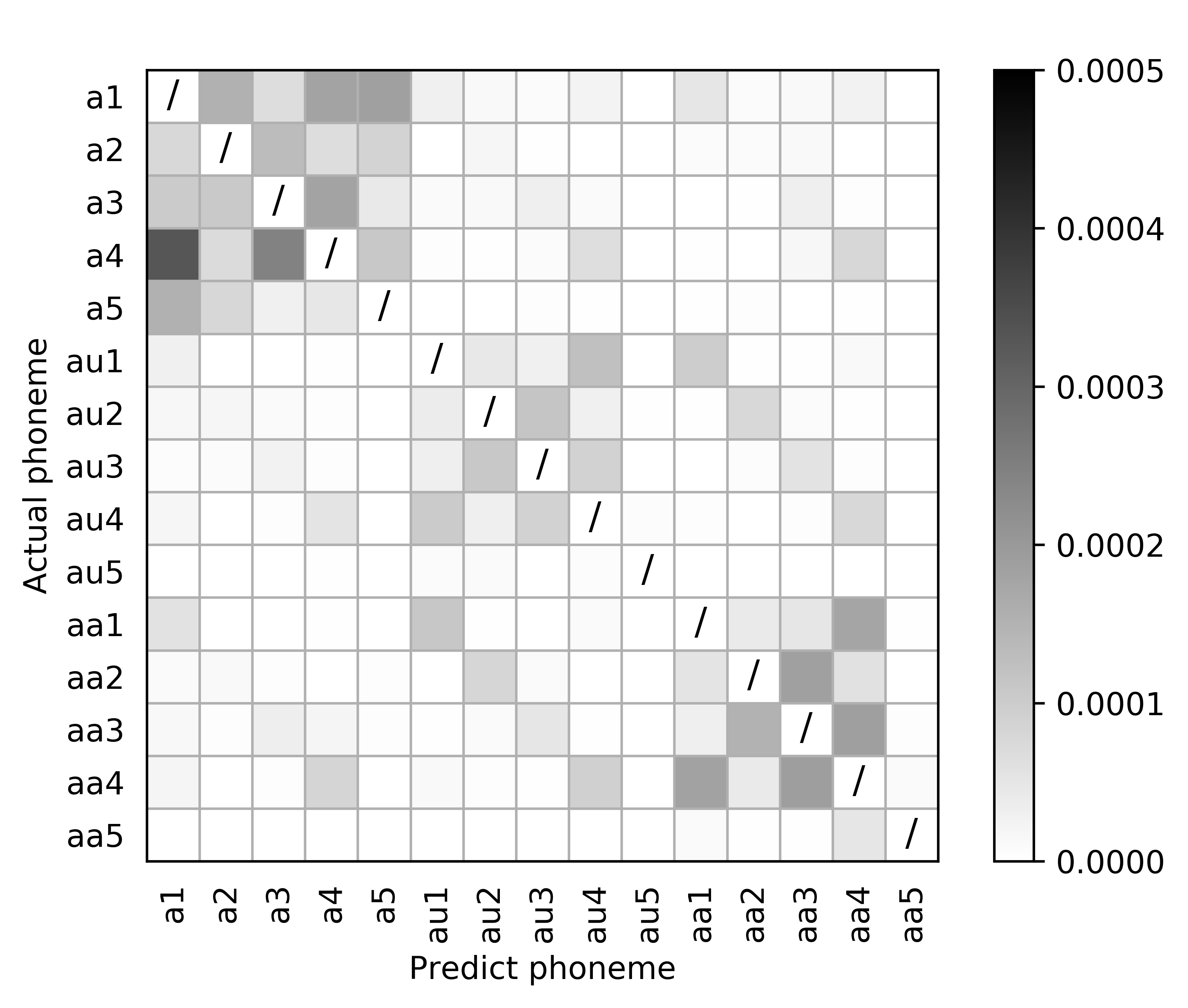}
		    \centerline{(a) AR set.}
		    \label{fig2}
		    \vspace{-13.0pt}
		\end{minipage}
	}
	\subfigure{
	    \begin{minipage}{6cm}
	    \centering
		\includegraphics[scale=0.5]{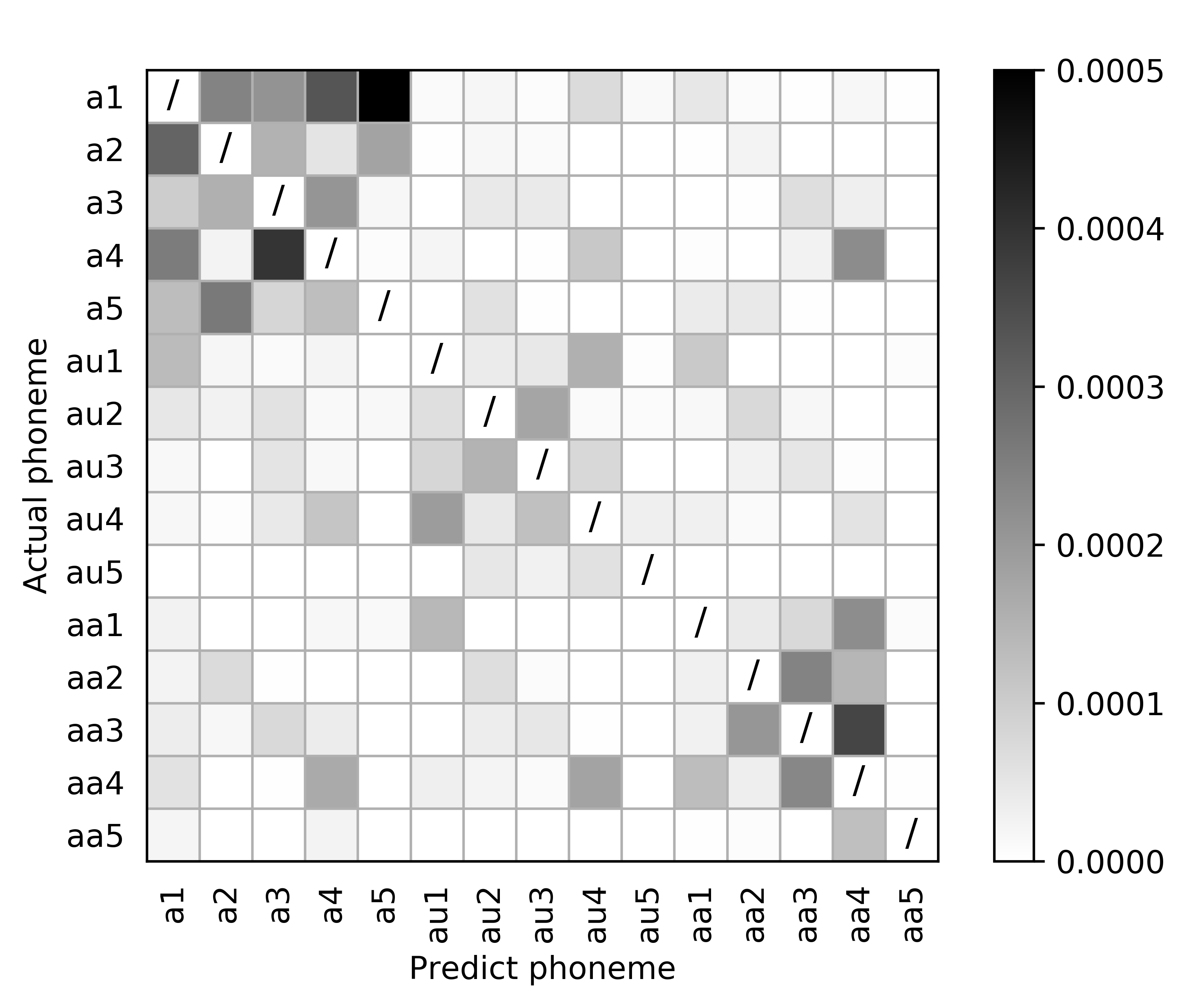}
	    \centerline{(b) CR set.}
	    \label{fig3}
	    \vspace{-13.0pt}
	    \end{minipage}
	}
	\caption{Confusion matrix of some phonemes on two sets. Note that diagonal is not filled for better showing the phoneme confusion.}
\end{figure}

First of all, we mix all the children and adult data together for training. Table~\ref{tab:11} reports the results of chain-based hybrid model on Kaldi,  transformer-based end-to-end system on ESPnet and CTC-CRF-based system on CAT. According to Table~\ref{tab:11}, the CER on CC is significantly worse than that of the other two sets. This indicates that children's conversational speech is more challenging to recognize. We further try simple model adaptation. Specifically, for the three systems, we merge children training data from C1, C2 and the same amount of adult data from Set A. Then we use the merged data to fine-tune the model B0, B1 and B2. As shown in Table~\ref{tab:11}, the CER of the chain model, the transformer model and the CTC-CRF model on the CC set decreases relatively by 2.5\%, 5.4\% and 0.9\% respectively. Meanwhile, the evaluation criteria $Score_{CER}$ decreases relatively by 2.3\%, 3.1\% and 1.3\% respectively. 
\begin{table}[!htb]
	\vspace{-10pt}
	\caption{Result comparison on development set in CER (\%).}
	\vspace{10pt}
	\label{tab:11}
	\centering
	\scalebox{0.8}{
		\begin{tabular}{l l c c c c }
			\toprule
			\multirow{2}{*}{\textbf{Exp ID}} & \multirow{2}{*}{\textbf{Model}} & \multicolumn{4}{c}{\textbf{CER (\%)}} \\
			\cline{3-6}	&& \textbf{AR} & \textbf{CR} & \textbf{CC} & $\textbf{Score}_{CER}$ \\ \hline 
			B0 & Chain model &  11.5 & 12.2 & 36.4 & 26.6\\ 
			B0A & + fine-tune & 12.0 & 12.1 & 35.5 & 26.0\\ 
			B1 & Transformer & 12.5 & 10.9 & 35.5 & 25.5\\  
			B1A & + fine-tune & 15.0 & 11.8 & 33.6 & 24.7\\
			B2 & CTC-CRF & 9.6 & 9.8 & 33.2 & 23.7\\  
			B2A & + fine-tune & 9.8 & 9.7 & 32.9 &
		    23.4\\
			\bottomrule
		\end{tabular}
	}
\end{table}
\\In order to further analyze the challenge of children's speech compared with adult speech, we use all the training data to train a phoneme classifier.  Fig.~\ref{fig2} shows the confusion matrix of some phonemes on the AR and CR development sets. It can be observed that compared with AR, phoneme discriminators are more likely to confuse similar phonemes on the CR set. This phenomenon is more obvious within \textit{\{a, aa, au\}} which are similar vowel phonemes.

\section{Challenge Rules and Schedule}	
\subsection{Rules}
All participants should adhere to the following rules to be eligible for the challenge.
\begin{enumerate}
	\item Data augmentation is allowed on the original training dataset, including, but not limited to, adding noise or reverberation, speed perturb and tone change.
	\item The use of evaluation datasets in any form of non-compliance is strictly prohibited, including but not limited to use the evaluation dataset to fine-tune or train the model.
	\item Multi-system fusion is not allowed.
	\item If the CER of the two systems on the evaluation dataset are the same, the system with lower computation complexity will be judged as the superior one.
	\item If forced alignment is used to obtain the frame-level classification label, the forced alignment model must be trained on the basis of the data allowed by the corresponding track.
	\item Shallow fusion is allowed for the end-to-end approaches, e.g., LAS, RNNT and Transformer, but the training data of the shallow fusion language model can only come from the transcripts of the training dataset.
	\item The right of final interpretation belongs to the organizer. In case of special circumstances, the organizer will coordinate the interpretation.
\end{enumerate}

\subsection{Timeline}	
\begin{itemize}
	\item \textbf{$August~14^{th}, 2020:$} Registration deadline, the due date for participants to join the Challenge.
	
	\item \textbf{$August~21^{st}, 2020:$} Training data release.
	
	\item \textbf{$September~30^{th}, 2020:$} Evaluation data release.
	
	\item \textbf{$October~10^{th}, 2020:$} Final submission deadline.
	
	\item \textbf{$October~25^{th}, 2020:$} Evaluation result and ranking release.
	
	\item \textbf{$November~8^{th}, 2020:$} Evaluation data release.
	
	\item \textbf{$January~19^{th}-22^{nd}, 2021:$} SLT2021 main workshop and challenge workshop.
\end{itemize}

\section{Conclusions}

In this paper, we introduce the necessity of IEEE SLT 2021 children speech recognition challenge (CSRC) and describe the datasets, tracks, rules and baselines for the challenge. With the released sizable open data and common testbed, we intend to advance children speech recognition and related techniques. After the challenge workshop, we plan to summarize the challenge results and introduce the best-performing approachs in another follow-up paper.

%% file: main.bbl
\begin{thebibliography}{10}

\bibitem{lecun2015deep}
Yann LeCun, Yoshua Bengio, and Geoffrey Hinton,
\newblock ``Deep learning,''
\newblock {\em nature}, vol. 521, no. 7553, pp. 436--444, 2015.

\bibitem{russell2007challenges}
Martin Russell and Shona D’Arcy,
\newblock ``Challenges for computer recognition of children’s speech,''
\newblock in {\em Workshop on SLT in Education}, 2007.

\bibitem{lee1997analysis}
Sungbok Lee, Alexandros Potamianos, and Shrikanth~S Narayanan,
\newblock ``Analysis of children’s speech. pitch and formant frequency,''
\newblock {\em Journal of the Acoustical Society of America}, vol. 101, no. 5,
  pp. 3194--3194, 1997.

\bibitem{lee1999acoustics}
Sungbok Lee, Alexandros Potamianos, and Shrikanth~S Narayanan,
\newblock ``Acoustics of children’s speech: Developmental changes of temporal
  and spectral parameters,''
\newblock {\em Journal of the Acoustical Society of America}, vol. 105, no. 3,
  pp. 1455--1468, 1999.

\bibitem{potamianos1997automatic}
Alexandros Potamianos, Shrikanth Narayanan, and Sungbok Lee,
\newblock ``Automatic speech recognition for children,''
\newblock in {\em Fifth European Conference on Speech Communication and
  Technology}, 1997.

\bibitem{das1998improvements}
Subrata Das, Don Nix, and Michael Picheny,
\newblock ``Improvements in children's speech recognition performance,''
\newblock in {\em IEEE International Conference on Acoustics, Speech and Signal
  Processing}, 1998, pp. 433--436.

\bibitem{Li2002An}
Qun Li and Martin~J. Russell,
\newblock ``An analysis of the causes of increased error rates in childrens
  speech recognition.,''
\newblock in {\em International Conference on Spoken Language Processing},
  2002, pp. 2337--2340.

\bibitem{gray2014child}
Sharmistha~S Gray, Daniel Willett, Jianhua Lu, Joel Pinto, Paul Maergner, and
  Nathan Bodenstab,
\newblock ``Child automatic speech recognition for {US} english: child
  interaction with living-room-electronic-devices.,''
\newblock in {\em Workshop on Child Computer Interaction}, 2014, pp. 21--26.

\bibitem{Giuliani2015LargeVC}
Diego Giuliani and Bagher BabaAli,
\newblock ``Large vocabulary children's speech recognition with {DNN}-{HMM} and
  {SGMM} acoustic modeling,''
\newblock in {\em INTERSPEECH}, 2015, pp. 1635--1639.

\bibitem{Cosi2015A}
Piero Cosi,
\newblock ``A {KALDI}-{DNN}-based {ASR} system for {Italian},''
\newblock in {\em International Joint Conference on Neural Networks}, 2015, pp.
  1--5.

\bibitem{7078563}
Romain Serizel and Diego Giuliani,
\newblock ``Vocal tract length normalisation approaches to {DNN}-based
  children's and adults' speech recognition,''
\newblock in {\em IEEE Spoken Language Technology Workshop}, 2014, pp.
  135--140.

\bibitem{Liao2015Large}
Hank Liao, Golan Pundak, Olivier Siohan, Melissa Carroll, Noah Coccaro, Qi-Ming
  Jiang, Tara~N Sainath, Andrew Senior, Fran{\c{c}}oise Beaufays, and Michiel
  Bacchiani,
\newblock ``Large vocabulary automatic speech recognition for children,''
\newblock 2015.

\bibitem{Qian2017Mismatched}
Mengjie Qian, Ian McLoughlin, Wu~Quo, and Lirong Dai,
\newblock ``Mismatched training data enhancement for automatic recognition of
  children's speech using {DNN}-{HMM},''
\newblock in {\em International Symposium on Chinese Spoken Language
  Processing}, 2016, pp. 1--5.

\bibitem{Fainberg2016Improving}
Joachim Fainberg, Peter Bell, Mike Lincoln, and Steve Renals,
\newblock ``Improving children's speech recognition through out-of-domain data
  augmentation,''
\newblock in {\em INTERSPEECH}, 2016, pp. 1598--1602.

\bibitem{inproceedings}
Yao Qian, Xinhao Wang, Keelan Evanini, and David Suendermann-Oeft,
\newblock ``Improving {DNN}-based automatic recognition of non-native children
  speech with adult speech,''
\newblock in {\em Workshop on Child Computer Interaction}, 2016, pp. 40--44.

\bibitem{Rong2017Transfer}
Rong Tong, Lei Wang, and Bin Ma,
\newblock ``Transfer learning for children's speech recognition,''
\newblock in {\em International Conference on Asian Language Processing}, 2017,
  pp. 36--39.

\bibitem{DBLP:journals/corr/abs-1809-09658}
Marco Matassoni, Roberto Gretter, Daniele Falavigna, and Diego Giuliani,
\newblock ``Non-native children speech recognition through transfer learning,''
\newblock in {\em IEEE International Conference on Acoustics, Speech and Signal
  Processing}, 2018, pp. 6229--6233.

\bibitem{Gurunath2018Transfer}
Prashanth Gurunath~Shivakumar and Panayiotis Georgiou,
\newblock ``Transfer learning from adult to children for speech recognition:
  Evaluation, analysis and recommendations,''
\newblock 2020, pp. 101077--101077.

\bibitem{Povey2016Purely}
Daniel Povey, Vijayaditya Peddinti, Daniel Galvez, Pegah Ghahremani, Vimal
  Manohar, Xingyu Na, Yiming Wang, and Sanjeev Khudanpur,
\newblock ``Purely sequence-trained neural networks for asr based on
  lattice-free mmi,''
\newblock in {\em INTERSPEECH}, 2016, pp. 2751--2755.

\bibitem{DBLP:journals/corr/VaswaniSPUJGKP17}
Ashish Vaswani, Noam Shazeer, Niki Parmar, Jakob Uszkoreit, Llion Jones,
  Aidan~N Gomez, {\L}ukasz Kaiser, and Illia Polosukhin,
\newblock ``Attention is all you need,''
\newblock in {\em Advances in neural information processing systems}, 2017, pp.
  5998--6008.

\bibitem{CAT_IS20}
Keyu An, Hongyu Xiang, and Zhijian Ou,
\newblock ``{CAT}: A {CTC-CRF} based {ASR} toolkit bridging the hybrid and the
  end-to-end approaches towards data efficiency and low latency,''
\newblock in {\em INTERSPEECH}, 2020.

\bibitem{povey2011kaldi}
Daniel Povey, Arnab Ghoshal, Gilles Boulianne, and \emph{et al},
\newblock ``The {KALDI} speech recognition toolkit,''
\newblock in {\em IEEE workshop on automatic speech recognition and
  understanding}, 2011.

\bibitem{graves2006connectionist}
Alex Graves, Santiago Fern{\'a}ndez, Faustino Gomez, and J{\"u}rgen
  Schmidhuber,
\newblock ``Connectionist temporal classification: labelling unsegmented
  sequence data with recurrent neural networks,''
\newblock in {\em Proceedings of the 23rd international conference on Machine
  learning}, 2006, pp. 369--376.

\bibitem{dong2018speech}
Linhao Dong, Shuang Xu, and Bo~Xu,
\newblock ``Speech-transformer: A no-recurrence sequence-to-sequence model for
  speech recognition,''
\newblock in {\em IEEE International Conference on Acoustics, Speech and Signal
  Processing}, 2018, pp. 5884--5888.

\bibitem{li2019speechtransformer}
Jie Li, Xiaorui Wang, Yan Li, et~al.,
\newblock ``The speechtransformer for large-scale mandarin chinese speech
  recognition,''
\newblock in {\em IEEE International Conference on Acoustics, Speech and Signal
  Processing}, 2019, pp. 7095--7099.

\bibitem{zeyer2019comparison}
Albert Zeyer, Parnia Bahar, Kazuki Irie, Ralf Schl{\"u}ter, and Hermann Ney,
\newblock ``A comparison of transformer and {LSTM} encoder decoder models for
  {ASR},''
\newblock in {\em IEEE Automatic Speech Recognition and Understanding
  Workshop}, 2019, pp. 8--15.

\bibitem{karita2019comparative}
Shigeki Karita, Nanxin Chen, Tomoki Hayashi, and \emph{et al},
\newblock ``A comparative study on transformer vs {RNN} in speech
  applications,''
\newblock in {\em IEEE Automatic Speech Recognition and Understanding
  Workshop}, 2019, pp. 449--456.

\bibitem{Hu2019Phoneme}
Ke~Hu, Antoine Bruguier, Tara~N Sainath, Rohit Prabhavalkar, and Golan Pundak,
\newblock ``Phoneme-based contextualization for cross-lingual speech
  recognition in end-to-end models,''
\newblock {\em arXiv preprint arXiv:1906.09292}, 2019.

\bibitem{Karita2019Improving}
Tomohiro Nakatani,
\newblock ``Improving transformer-based end-to-end speech recognition with
  connectionist temporal classification and language model integration,''
\newblock in {\em INTERSPEECH}, 2019, pp. 1408--1412.

\bibitem{CRF_IC20}
Hongyu Xiang and Zhijian Ou,
\newblock ``{CRF}-based single-stage acoustic modeling with {CTC} topology,''
\newblock in {\em Proc. ICASSP}, 2019, pp. 5676--5680.

\bibitem{Watanabe2018ESPnet}
Shinji Watanabe, Takaaki Hori, Shigeki Karita, Tomoki Hayashi, Jiro Nishitoba,
  Yuya Unno, Nelson Enrique~Yalta Soplin, Jahn Heymann, Matthew Wiesner, Nanxin
  Chen, et~al.,
\newblock ``{ESPNET}: End-to-end speech processing toolkit,''
\newblock {\em arXiv preprint arXiv:1804.00015}, 2018.

\end{thebibliography}
